\documentclass{aa}
\usepackage{graphicx}
\usepackage{txfonts}
\usepackage{psfig}
\usepackage{psfrag}
\usepackage{latexsym}
\usepackage{times,amssymb,lscape,verbatim}
\usepackage{natbib}
\usepackage[mathcal]{euscript}
\usepackage{mathrsfs}

\begin{document}
	\title{Large grains in disks around young stars: ATCA observations of WW~Cha, RU~Lup, and CS~Cha}
   \titlerunning{ATCA observations of WW~Cha, RU~Lup, CS~Cha}

   \author{D. Lommen\inst{1}
   	\and S. T. Maddison\inst{2,3}
	\and C. M. Wright\inst{4}	
	\and E. F. van Dishoeck\inst{1,5}
	\and D. J. Wilner\inst{6}
	\and T. L. Bourke\inst{6}
	     }

   \offprints{Dave Lommen, \\ \email{dave@strw.leidenuniv.nl}}

   \institute{Leiden Observatory, Leiden University, P.O. Box 9513, 2300 RA Leiden, The Netherlands
   	\and Centre for Astrophysics and Supercomputing, Swinburne University of Technology, PO Box 218, Hawthorn, VIC 3122, Australia
	\and Centre de Recherche Astrophysique de Lyon, \'Ecole Normale Sup\'erieure de Lyon, 46 all\'ee d'Italie, F-69364 Lyon cedex 07, France
	\and School of Physical, Environmental and Mathematical Sciences, UNSW@ADFA, Canberra ACT 2600, Australia
	\and Max-Planck-Institut f\"{u}r Extraterrestrische Physik (MPE), Giessenbachstr. 1, 85748 Garching, Germany
	\and Harvard-Smithsonian Center for Astrophysics, 60 Garden Street, 02138 Cambridge, MA, USA
 }

   \date{Received ...; accepted ...}

 \abstract 
  {Grains in disks around young stars grow from interstellar submicron sizes to planetesimals, up to thousands of 
  kilometres in size, over the course of several Myr. Thermal emission of large grains or pebbles can be best observed at 
  centimetre wavelengths. However, other emission mechanisms can contribute, most notably free-free emission from stellar 
  winds and chromospheric activity.}
  {We aim to determine the mechanisms of centimetre emission for three T~Tauri stars.
  WW~Cha and RU~Lup were recently found to have grain growth at least up to 
  millimetre sizes in their circumstellar disks, based on millimetre data up to 3.3~mm. CS~Cha has   similar indications for 
  grain growth in its circumbinary disk.}
  {The T~Tauri stars WW~Cha and RU~Lup were monitored over the course of several years at millimetre and centimetre wavelengths, using
  the Australia Telescope Compact Array (ATCA).  
  The new ATCA 7~mm system was also used to observe CS~Cha at 7~mm.}
  { WW~Cha was detected on several occasions at 7 and 16~mm.  We obtained one detection of WW~Cha at 3.5~cm and upper limits only 
  for 6.3~cm.  The emission at 16~mm was stable over periods of days, months and years, whereas the emission at 3.5~cm is found to be variable.
  A second young stellar object, Ced~112~IRS~4, was found in the field of WW~Cha at 16~mm.
  RU~Lup was detected at 7~mm. It was observed at 16~mm three times and at 3 and 6~cm four times and found to be variable in all three 
  wavebands. CS~Cha was detected at 7~mm, but the signal-to-noise was not high enough to resolve the gap in the circumbinary disk. 
  The typical resolution of the 7 and 16~mm observations were 5--10 arcsec with rms $\sim0.2$ mJy. }
  {The emission at 3, 7 and 16~mm for WW~Cha is well explained by thermal emission from millimetre and centimetre-sized 
  ``pebbles.'' The cm spectral index between 3.5 and 6.3~cm is consistent with the emission from an optically-thick ionised 
  wind, although the high variability of the cm emission points to a non-thermal contribution. The spectral energy  distributions of both 
  RU~Lup and CS~Cha from 1 to 7~mm are consistent with thermal emission from mm-sized grains.  The variability of the 
  longer-wavelength emission for RU~Lup and the negative spectral index suggests non-thermal emission, arising from an 
  optically-thin plasma.}


   \maketitle
%

\section{Introduction}\label{sect: introduction}

	Planet formation takes place in the disks around young stellar objects (YSOs), where submicron-sized grains
	 have to grow to planet size in about 10~Myr or less 
	\citep{haisch:2001, carpenter:2005, setiawan:2008}. The very first step, where the grains grow from submicron 
	size to several microns, may be traced by the 10-$\mu$m silicate feature. Silicates of a few microns in size give 
	considerably flatter features than those of submicron sizes \citep[e.g.][]{przygodda:2003}. However, the strength 
	and shape of the 10-$\mu$m feature may also be influenced by the crystallinity of the grains 
	\citep[e.g.,][]{meeus:2003}, or by their porosity \citep[][]{voshchinnikov:2008}. Furthermore, the 10-$\mu$m
	silicate feature only probes the surface layers of the circumstellar disks.
	
	Growth up to millimetre (mm) sizes can be more readily observed by looking at the mm slope in the spectral energy 
	distribution (SED). From the slope $\alpha$, where $F_\nu \propto \nu^{\, \alpha}$, the opacity index $\beta$ in the mm 
	regime, where $\kappa_\nu \propto \nu^{\, \beta}$, can be obtained through
	\begin{equation}
		\label{eq: beta}
		\beta \approx (\alpha - 2)(1 + \Delta),
	\end{equation}
	where $\Delta$ is the ratio of optically thick to optically thin emission from the disk 
	\citep{beckwith:1990, beckwith:1991, rodmann:2006}. An opacity index $\beta \approx 1$ indicates that grains have 
	grown at least up to mm sizes, irrespective of their exact physical structure \citep{draine:2006}. Indeed, about 30 
	sources with grains up to at least mm sizes have been found over the past several years 
	\citep[][]{natta:2004,rodmann:2006,andrews:2007,lommen:2007}.
	
	The next step, growth to centimetre (cm) sizes and beyond, is much harder to observe directly. The reason for this 
	is that at cm wavelengths, where these ``pebbles'' can be observed, the emission is several orders of magnitudes
	weaker than at mm wavelengths. Furthermore, other processes such as chromospheric activity 
	\citep[][]{forbrich:2007}, gyrosynchrotron emission \citep[][]{andre_etal:1988}, electron-cyclotron maser emission 
	\citep[][]{dulk:1985,smith:2003} or an ionised wind \citep[][]{girart:2004} may also contribute significantly to the cm
	emission.  A particularly interesting source of variable cm emission may be that from (interacting) helmet streamers
	\citep{massi:2006,massi:2008}. To rule out these other cm emission mechanisms, the sources have to be monitored over days, 
	months, and years, to ascertain that the cm emission is optically thin and not variable, and ideally be spatially resolved. 
	So far, TW~Hya is the only T~Tauri star that has been monitored over sufficient time periods to safely
        characterise the cm emission as thermal emission from pebbles \citep{wilner:2005}. [\citet{natta:2004} and 
        \citet{alonso-albi:2008} also claim 
        thermal emission from cm-sized grains though their observations have not yet been monitored over an extended period of time.]
	
	\citet{lommen:2007} observed T~Tauri stars in the southern constellations Lupus and Chamaeleon with the Australia 
	Telescope Compact Array (ATCA\footnote{The Australia Telescope Compact Array is part of the Australia Telescope 
	which is funded by the Commonwealth of Australia for operation as a National Facility managed by CSIRO.}) at 3.3~mm, 
	and found three sources that were resolved and had an opacity index $\beta < 1.0$. We chose the two brightest of 
	these, WW~Cha and RU~Lup, for follow-up observations with the ATCA at 7~mm through 6~cm, the results of which are 
	presented in this paper. The source CS~Cha was not resolved at 3~mm but found to have an opacity index of 
	$\beta = 1.0 \pm 0.6$ by 
	\citet{lommen:2007} and was added for high-resolution observations at 7~mm. 

	WW~Cha is located in a reflection nebula in the Ced~112 region of the Chamaeleon~I molecular cloud, and is thought 
	to drive the highly collimated jets HH~915 and HH~931 \citep{bally:2006, wang:2006}. \citet{wang:2006} claim that 
	two near-infrared molecular-hydrogen emission knots detected by \citet{gomez:2004} on the opposite side of WW~Cha 
	may be the counterparts of HH~915. WW~Cha has a relatively weak 10-$\mu$m amorphous-silicate feature 
	\citep{przygodda:2003}, indicating that the surface layers are dominated by micron-sized grains. No clear 
	crystalline features are detected in the 10-$\mu$m region. \citet{reipurth:1996} placed the bright mm source
	Cha-MMS2 about $9\arcsec$ from IRAS~11083-7618, which is the infared counterpart to the T~Tauri star WW~Cha.
	However, given the similar fluxes found from SEST observations by both \citet{henning:1993} and \citet{reipurth:1996},
	the lack of other mm sources within several tens of arcseconds, and the fairly large beam used for these and the
	IRAS observations, we claim that Cha-MMS2 and WW~Cha are one and the same source. Another T~Tauri star, Ced~112~IRS~4,
	is located about $40\arcsec$ to the north from WW~Cha. 

	RU~Lup is a very active and well-studied T~Tauri star, showing variability in the optical, UV and X-ray bands 
	\citep[e.g.][]{lamzin:1996, stempels:2002, herczeg:2005, robrade:2007}. The mass accretion rate onto the central 
	star is found to be relatively large, at $\sim 10^{-7}$~M$_\odot$~yr$^{-1}$ \citep{lamzin:1996, podio:2007}. 
	\citet{stempels:2002} studied the variations in the radial velocity in RU~Lup, which they found to be periodic with 
	a period of 3.71~days. \citeauthor{stempels:2002} attribute this periodic variability to long-lived star spots, and 
	find the solution of a substellar companion unlikely. Olofsson et al. (2008, in prep.) show a very strong and boxy 10-$\mu$m 
	feature for RU~Lup, indicating the presence of submicron-sized particles in the disk photosphere. Contributions from 
	crystalline silicates are detected between 20 and 35~$\mu$m \citep{kessler-silacci:2006}.

        CS~Cha was classified as a so-called transitional disk by  \citet{espaillat:2007a}.  The SEDs of transitional disks 
	show a lack of infrared emission, indicating a deficit of warm dust close to the star.  SEDs of transitional 
	disks are well fit by models that include an inner hole, suggesting that the disks are in a  transitional stage. The loss of  		warm dust can be explained by photo-evaporation, by dust growing to larger sizes 
	and effectively moving the flux to longer wavelengths in the SED, and/or by the presence of an unseen 
	planet that sweeps up material in the inner disk.  However, CS~Cha was recently found to be a binary 
	\citep{gunther:2007}, and so is now classified as a circumbinary disk rather than a transitional disk.
	Basic parameters of the 	
	sources WW~Cha, RU~Lup, and CS~Cha are presented in Table~\ref{tab: source list}.
	\begin{table*}
	 \caption[]{Source list of sources observed with the ATCA.}
	 \label{tab: source list}
	 \centering
	  \begin{tabular}{lllcclll}
	   \hline
	   \hline
	   Source          	& Cloud		& Distance$^\mathrm{a}$ & Age$^\mathrm{b}$ 	& Luminosity$^\mathrm{b}$	& Mass$^\mathrm{b}$	& Spectral 		& Position$^\mathrm{d}$		\\
	                   	&		& (pc)			& (Myr)			& (L$_\odot$)			& (M$_\odot$)		& type$^\mathrm{c}$	&				\\
	   \hline
	   WW Cha      & Cha I         & $160\pm15$	       & 0.4--0.8	       & 2.2			       & 0.6--0.8	       & K5		       & (11:10:00.7, -76:34:59)       \\
	   RU Lup      & Lup II        & $140\pm20$	       & 0.04--0.5	       & 2.1			       & 2.0--2.8	       & K7-M0  	       & (15:56:42.3, -37:49:15.47)    \\
	   CS Cha      & Cha I         & $160\pm15$	       & 2--3		       & 1.3			       & 0.9--1.2	       & K4		       & (11:02:25.1, -77:33:35.95)    \\
	   \hline
	  \end{tabular}
	 \begin{list}{}{}
	  \item[$^\mathrm{a}$]     Distances from \citet{whittet:1997} (WW~Cha and CS~Cha) and \citet{hughes:1993} (RU~Lup). 
	  \item[$^\mathrm{b}$]     Ages, luminosities and masses from \citet{hughes:1994} (RU~Lup) and \citet{lawson:1996} (WW~Cha and CS~Cha).
	  \item[$^\mathrm{c}$]     Spectral types from \citet{gauvin:1992} (WW~Cha and CS~Cha) and \citet{hughes:1994} (RU~Lup). 
	  \item[$^\mathrm{d}$]     RA and dec positions (J2000) from SIMBAD. 
	 \end{list} 
	\end{table*}
		
	We here present observations of WW~Cha and RU~Lup, taken with the ATCA at wavelengths ranging from 
	7~mm to 6.3~cm, 
	to determine their cm emission mechanisms.  We also present ATCA 7~mm observations of the 
	binary CS~Cha, with the aim to obtain a longer-wavelength flux point and resolve the hole in the dust disk, which has
      a diameter of $\sim 85$~AU \citep{espaillat:2007a}, at mm wavelengths. The observations are described in 
	Sect.~\ref{sect: observations}, with the basic results presented in Sect.~\ref{sect: results} and further discussed 
	in Sect.~\ref{sect: discussion}.  Conclusions are presented in Sect.~\ref{sec:concs}.


\section{Observations}\label{sect: observations}

	We present continuum observations of WW~Cha, RU~Lup, and CS~Cha, observed with the ATCA over the period 2006-2008 at 
	7 and 16~mm and at 3.5 and 6.3~cm.  The observations are listed in Table~\ref{tab: observations}.  ATCA is an array of  $6\times 22$~m 
	antennas, with antenna 6 (CA06) fixed at 6~km.  The observations were carried out in double sideband, where each sideband had a 
	bandwidth of 128~MHz. CA06 was only included in the reduction when the array was in an extended configuration 
	(1.5A, 1.5B or 6D -- where the number in these array configuration is approximately the longest baseline in kilometres).	 
	The data were calibrated and imaged using the MIRIAD package \citep{sault:1995}.

	The complex gain calibration was done on the calibrators QSO~B1057-797 (for WW~Cha and CS~Cha) and QSO~B1622-297 
	(for RU~Lup), both of which are within 10~degrees of the science targets. The complex gains for RU~Lup were twice 
	calibrated using different sources: on QSO~B1600-44 (on 9 June 2007) and on QSO~B1622-310 (on 4 November 2006).

	Normally the absolute flux calibration was done on Uranus at 7~mm and on ATCA's primary flux calibrator QSO~B1934-638 
	at cm wavelengths. QSO~B1934-638 is stable at cm wavelengths, and its 
	flux as a function of frequency is well known. On two occasions the absolute flux 
	calibration was done on Mars. The baselines on which Mars was resolved out were not used in the flux calibration. 
	The absolute flux calibration was done on QSO~B1057-797 and QSO~B1921-293 on four occasions.  See 
	Table~\ref{tab: observations} for details.
	
	Using QSO~B1057-797 or QSO~B1921-293 as an absolute flux calibrator does require some care. However, in some cases we have data 
	very close in time to when the Observatory has published fluxes (e.g., the previous or next day for QSO~B1057-797 in October 2007),
	and the long-term variability of QSO~B1057-797 is at most a factor of 2, and that of QSO~B1921-293 likely much smaller than that
	(see the ATCA calibrator pages at {\tt {\small http://www.narrabri.atnf.csiro.au/calibrators/}}). 
	In this work, we assume an uncertainty in the calibrated flux of about 15\%, unless stated otherwise.
		
	Note that for all of our observations the phase centre is offset from the source by one or two synthesised beamwidths in right ascension. 
	This was done to avoid any artefacts at the centre of the field.

\subsection{WW~Cha}

	Observations of  WW~Cha at 7~mm were carried twice in compact configurations. On both days
	the observations did not include a planet for the flux calibrator and the fluxes were calibrated on the gain calibrator 
	QSO~B1057-797, resulting in an estimated uncertainty in the absolute flux calibration of 15\% in  
	October 2007 and 30\% in March 2008.
	We conducted six observations of WW~Cha at 16~mm in the period May 2006-March 2008, all
	in compact configurations except in November 2007 when an extended array was used.
	Without a dedicated flux calibrator on 31 March 2008, the estimated uncertainty for this track is 30\%.
	WW~Cha was observed at 3 and 6~cm twice.
	
\subsection{RU~Lup}

	RU~Lup was observed at 7~mm  just once, while  three 16~mm observations were conducted, once in an extended 
	configuration. One of the four observations made at 3 and 6~cm was in an extended array configuration. However a large fraction of 
	the data on the longest baselines had to be flagged, losing the advantage of the extended configuration. Fluxes are thought to be 
	calibrated to an uncertainty of $\sim$10\% in October 2006.

\subsection{CS~Cha}

	Because the initial aim for CS~Cha was to resolve the hole in the circumbinary disk, this source was only observed while the ATCA 
	was in an extended configuration. CS~Cha was observed four times at 7~mm in 2008, including a full track
	on 30 June,  though poor weather rendered these data unusable.

	\begin{table*}
		\caption[]{Overview of the observations.}\label{tab: observations}
		\centering
		\begin{tabular}{lccccl}
			\hline
			\hline
			Obs. date				& Wavelengths 	& Configuration	& Flux calibrator	& Integration$^\mathrm{a}$ & Notes	\\
								& (mm)              	& 	        		&				& (hours)				& (weather, array \& data)\\
			\hline
			\multicolumn{6}{c}{WW~Cha, ATCA 7 mm band}																\\
			\hline
			5 October 2007				& 7.0, 7.3		& H75C		& 1057-797		& 1.35	& CA02 offline			\\
			31 March 2008				& 7.0, 7.3		& H168		& 1057-797		& 1.99		&		\\
			\hline	
			\multicolumn{6}{c}{WW~Cha, ATCA 12 mm band}																\\
			\hline
			8 May 2006$^\mathrm{b}$			& 16.1, 16.2		& H214C		& Mars			& 0.80 &excellent conditions; CA03$+$CA04 offline				\\
			13 October 2006				& 16.1, 16.2		& H214C		& 1934-638		& 1.31	& very good conditions; noisy 16.1~mm data			\\
			18 October 2006				& 16.1, 16.2		& EW352		& 1934-638		& 3.32	& noisy 16.1~mm data\\
			24 October 2007 			& 15.4, 16.2		& H214C		& 1934-638		& 3.15		& conditions deteriorated over scan		\\
			2 November 2007				& 15.4, 16.2		& 1.5A		& 1934-638		& 8.48	&			\\
			31 March 2008				& 15.4, 16.2		& H168		& 1057-797		& 1.33		&		\\
			\hline
			\multicolumn{6}{c}{WW~Cha, ATCA 3+6 cm bands}																	\\
			\hline
			18 October 2006				& 34.7, 62.5		& EW352		& 1934-638		& 3.95	& a lot of data flagged			\\
			9 June 2007				& 34.7, 62.5		& EW352		& 1934-638		& 5.19		&		\\
			\hline\hline
			\multicolumn{6}{c}{RU~Lup, ATCA 7 mm band}																\\
			\hline
			6 October 2007				& 7.0, 7.3		& H75C		& Uranus			& 0.99	& CA04 offline		\\
			\hline
			\multicolumn{6}{c}{RU~Lup, ATCA 12 mm band}																\\
			\hline
			11 October 2006				& 16.1, 16.2		& H214C		& 1934-638		& 3.66	& very good conditions\\
			24 October 2007				& 16.1, 16.2		& H214C		& 1934-638		& 2.49	& fair conditions	\\
			4 November 2007				& 15.4, 16.2		& 1.5A		& 1934-638		& 3.79	& fair conditions	\\
			\hline
			\multicolumn{6}{c}{RU~Lup, ATCA 3+6 cm bands}																	\\
			\hline
			12 October 2006				& 34.7, 62.5		& H214C		& 1934-638		& 3.53	& very good conditions\\
			13 October 2006 			& 34.7, 62.5		& H214C		& 1934-638		& 1.70		&very good conditions\\
			9 June 2007				& 34.7, 62.5		& EW352		& 1934-638		& 3.32		& good conditions		\\
			4 November 2007				& 34.7, 62.5		& 1.5A		& 1934-638		& 3.48	& a lot of data flagged			\\
			\hline\hline
			\multicolumn{6}{c}{CS~Cha, ATCA 7 mm band}																\\
			\hline
			26 April 2008				& 7.0, 7.3		& 6A		& 1921-293		& 6.89			&fair conditions\\
			30 June 2008				& 6.7, 7.0		& 1.5B		& Mars			& 4.85		&poor conditions\\
			5 July 2008				& 6.7, 7.0		& 1.5B		& Uranus		& 0.72		&very good conditions\\
			6 July 2008				& 6.7, 7.0		& 1.5B		& Uranus		& 1.71		&fair conditions		\\
			\hline
		\end{tabular}
		\begin{list}{}{}
			\item[$^\mathrm{a}$]  Total time spent on the science target, before flagging and calibration of data.
			\item[$^\mathrm{b}$]  Three antennas only.
		\end{list}
	\end{table*}

\section{Results}\label{sect: results}
	
\subsection{WW~Cha}

	Table~\ref{tab: results WW Cha} summarises the fluxes on the different dates at the various wavelengths. 
        Note that  in the Table and below the uncertainties do not include the uncertainties from the absolute flux calibration as given in the previous section.
        Continuum fluxes 
	are from point-source fits in the ($u, v$) plane, where the source was detected at 3$\sigma$ or better. 
	For the cases where WW~Cha was not detected to at least 3$\sigma$, an upper limit of three times the root mean 
	square of the noise is quoted.
	
	\begin{table*}
		\caption[]{Overview of the results for WW~Cha.}\label{tab: results WW Cha}
		\centering
		\begin{tabular}{lcccc}
			\hline
			\hline
			Obs. date	& Wavelength	& Flux$^\mathrm{a}$	& RMS$^\mathrm{a}$	& Beam size$^\mathrm{b}$	\\
					& (mm) 		& (mJy)			& (mJy/beam)		& (arcsec)			\\
			\hline
			\multicolumn{5}{c}{ATCA 7 mm band}																		\\
			\hline
			5 Oct 2007	& 7.0		& $5.41\pm0.32$			& $0.313$	& $11 \times 11$		\\
			5 Oct 2007	& 7.3		& $3.93\pm0.29$			& $0.350$	& $11 \times 11$		\\
			31 Mar 2008	& 7.0		& $5.10\pm0.19$			& $0.231$	& $5.8 \times 5.3$		\\
			31 Mar 2008	& 7.3		& $5.19\pm0.17$			& $0.194$	& $6.1 \times 5.6$		\\
			\hline	
			\multicolumn{5}{c}{ATCA 12 mm band}																					\\
			\hline
			8 May 2006	& 16.1		& $1.16\pm0.31$			& $0.329$	& $23 \times 6$			\\
			8 May 2006	& 16.2		& $1.04\pm0.30$			& $0.158$	& $23 \times 6$			\\
			13 Oct 2006	& 16.1		& $<0.921$			& $0.307$	& $25 \times 9$			\\
			13 Oct 2006	& 16.2		& $1.08\pm0.21$			& $0.217$	& $26 \times 9$			\\
			18 Oct 2006	& 16.1		& $0.81\pm0.20$			& $0.222$	& $38 \times 7$			\\
			18 Oct 2006	& 16.2		& $1.23\pm0.19$			& $0.110$	& $39 \times 7$			\\
			24 Oct 2007 	& 15.4		& $1.01\pm0.16$			& $0.209$	& $12 \times 8$			\\
			24 Oct 2007 	& 16.2		& $0.95\pm0.13$			& $0.181$	& $13 \times 9$			\\
			2 Nov 2007	& 15.4		& $0.55\pm0.20^\mathrm{c}$	& $0.125$	& $1.9 \times 1.3$		\\
			2 Nov 2007	& 16.2		& $0.48\pm0.13^\mathrm{c}$	& $0.089$	& $2.0 \times 1.4$		\\
			31 Mar 2008	 & 15.4		& $0.88\pm0.12$			& $0.232$	& $14 \times 13$		\\
			31 Mar 2008	 & 16.2		& $1.07\pm0.10$			& $0.227$	& $15 \times 13$		\\
			\hline
			\multicolumn{5}{c}{ATCA 3 cm band}																					\\
			\hline
			18 Oct 2006	& 34.7		& $<0.222$			& $0.074$	& $71 \times 20$		\\
			9 June 2007	& 34.7		& $0.63\pm0.06$			& $0.076$	& $53 \times 16$		\\
			\hline
			\multicolumn{5}{c}{ATCA 6 cm band}																					\\
			\hline
			18 Oct 2006	& 62.5		& $<0.202$			& $0.067$	& $125 \times 32$		\\
			9 June 2007	& 62.5		& $<0.399$			& $0.133$	& $91 \times 26$		\\
			\hline
		\end{tabular}
		\begin{list}{}{}
			\item[$^\mathrm{a}$] Continuum fluxes are from point-source fits in the ($u, v$) plane. If the source was not detected at 
				3$\sigma$, a 3$\sigma$ upper limit is quoted. RMS calculated from the cleaned image.
			\item[$^\mathrm{b}$] Restored beam, using natural weighting. Note that the exact beam size depends
				on frequency and location of the source in the sky, as well as on the time of the observations,
				over which time range the data were taken, and exactly which data/baselines were included.
			\item[$^\mathrm{c}$] The 16~mm values from the 2 November 2007 data were obtained with antenna 6 included, causing the
				significantly lower point-source flux. Note that a Gaussian fit recovers the full flux, indicating
				that the source is probably extended.
		\end{list}
	\end{table*}

\begin{figure}
                \centering
                \includegraphics[width=\columnwidth]{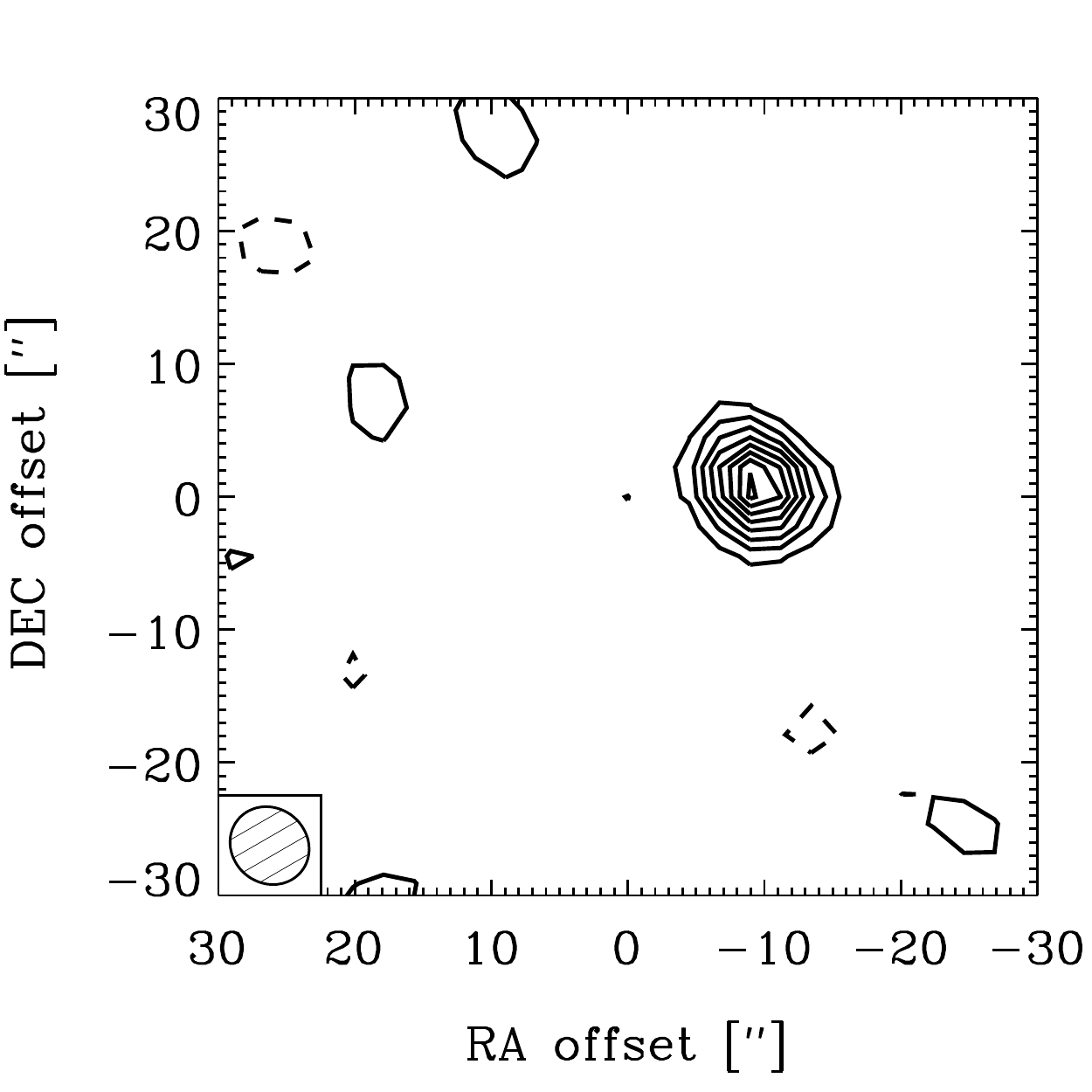}
		\caption[]{Image of WW~Cha, observed at 7.3~mm on 31 March 2008. The offsets are 
			with respect to the phase centre, which is located at 11:10:02.7, -76:34:59.0 (J2000); 
			peak emission coincides with the position of the optical star. The contours 
			are at 2, 4, 6, ... times the rms of 0.20~mJy/beam; negative contours are dashed.}
                \label{fig: wwcha at 7 mm}
        \end{figure}

	WW~Cha was observed with the ATCA at 3.3~mm in August 2005 and reported to have a point-source flux of 25.9~mJy by \citet{lommen:2007}. 
	At 7~mm, the source was detected in both sidebands on both occasions at which it was observed, and 
	the fluxes in the two sidebands were consistent from one date to the other to within the uncertainties of
	the absolute flux calibrations.  The fluxes at the different frequencies will be treated as separate 
	observations in the analysis. 
	Figure~\ref{fig: wwcha at 7 mm} shows WW~Cha at 7.3~mm. 
	
\begin{figure}
		\centering
		\includegraphics[angle=270,width=\columnwidth]{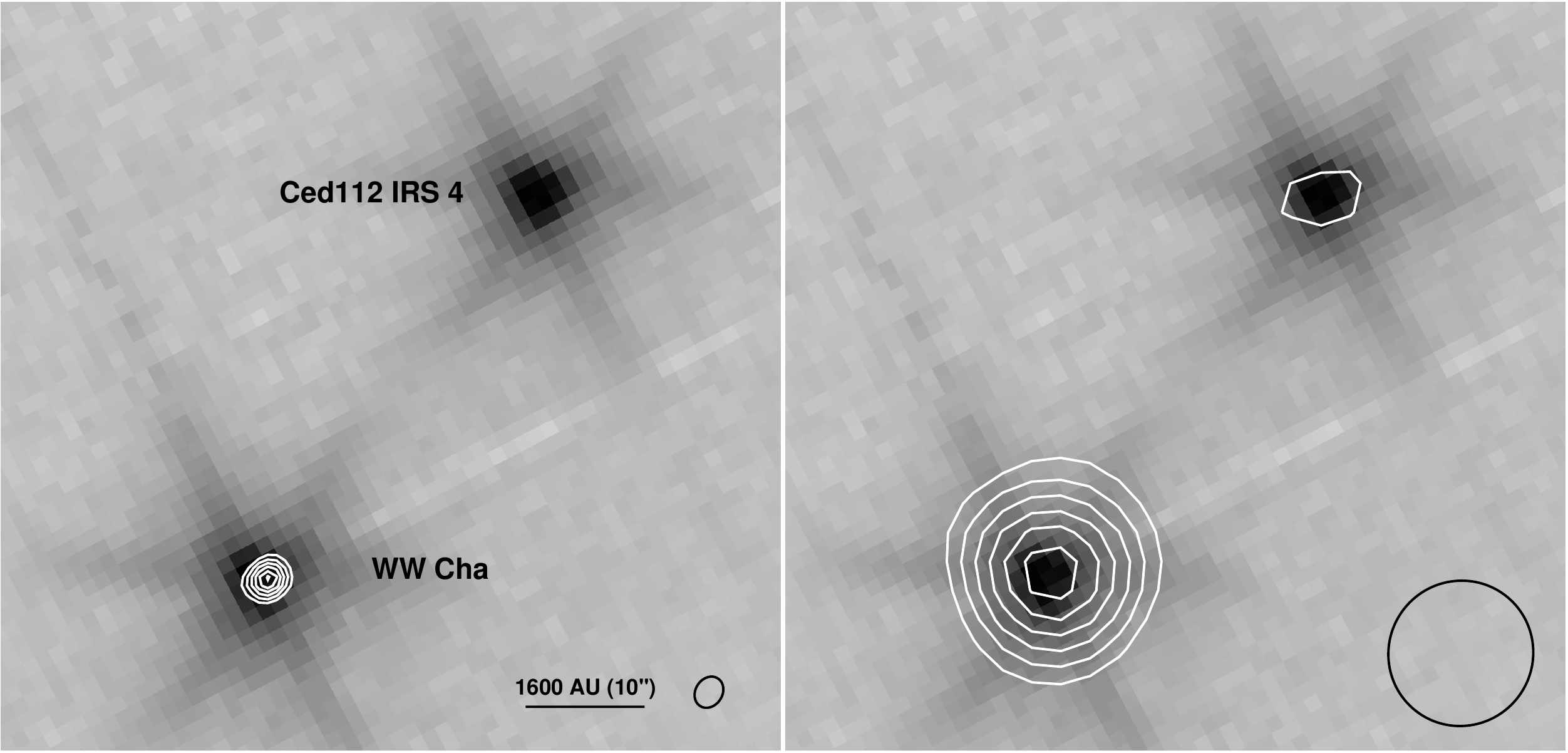}
		\caption[]{The contours show the emission towards WW~Cha, observed at 3~mm (left) and 16~mm (right), overplotted on
			{\it Spitzer Space Telescope} InfraRed Array Camera observations at 3.6~$\mu$m (greyscale). Contours are 4, 
			6, 8, ... times the rms (2.0~mJy/beam at 3~mm and 0.07~mJy/beam at 16~mm) and the size of the synthesised
			beams at the respective wavelengths is plotted in the lower right corner. The observations at 16~mm (from
			combined data of 8 May 2006, 13 October 2006, and 31 March 2008) also show a 4$\sigma$ detection of the
			young stellar object Ced~112~IRS~4.}
		\label{fig: IRAC}
	\end{figure}

\begin{figure}
		\centering
		\includegraphics[width=\columnwidth]{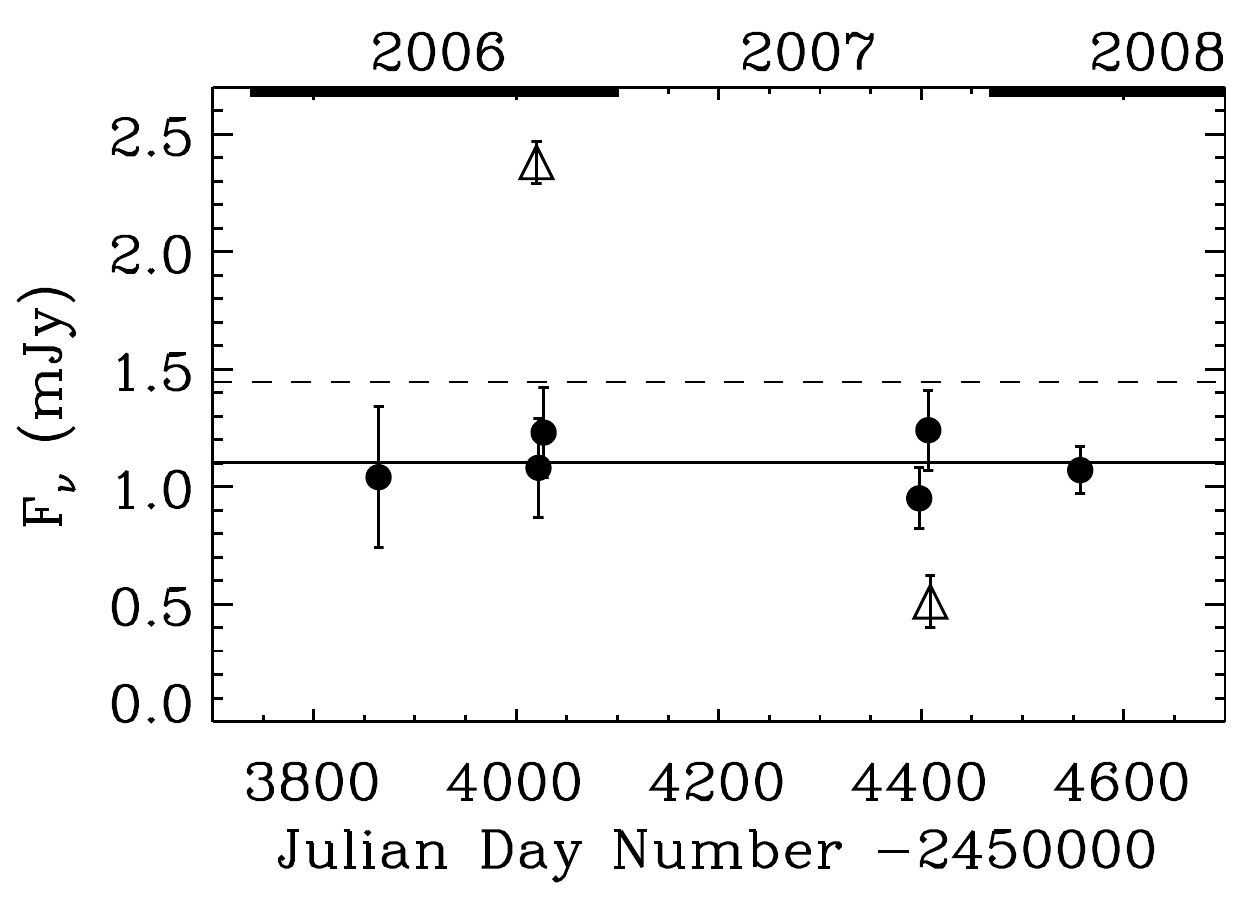}
		\caption[]{Point-source fits in the ($u, v$) plane for WW~Cha (dots) and RU~Lup (triangles) at 16.2~mm. The solid line
			shows the mean value of $1.10 \pm 0.48$~mJy for WW~Cha; the dashed line shows the mean value of $1.45 \pm 0.14$~mJy for
			RU~Lup. The thick lines at the top correspond to the years 2006 and 2008, for reference.
			Note that the upper limit of 24 October 2007 is omitted to prevent this figure from becoming too cluttered;
			the variability of RU~Lup at 16~mm is already clear from the two data points in the figure.}
		\label{fig: 18496 MHz}
	\end{figure}

	WW~Cha was observed at 16~mm six times in the period May 2006--March 2008. It was detected in both sidebands each time it was 	
	observed except on 13 October 2006, when an upper limit was obtained at 16.1~mm. See Fig.~\ref{fig: IRAC} for a comparison of 
	the 3~mm and 16~mm detections.
	For each frequency, the flux is constant over the different observations to within the uncertainties, as demonstrated by 
	Fig.~\ref{fig: 18496 MHz}, which shows the 
	results from point-source fits in the ($u, v$) plane at 16.2~mm, where CA06 was not included in the  2 November 2007 data  
	for consistency. Taking the arithmetic mean of the fluxes and quadratic mean of the uncertainties given in 
	Table~\ref{tab: results WW Cha}, average 
	point-source fluxes of $0.81\pm0.16$, $0.99\pm0.26$ and $0.98\pm0.19$~mJy at 15.4, 16.1 and 16.2~mm respectively were obtained. 

	WW~Cha was observed at 3.5 and 6.3~cm twice. A $3\sigma$ upper limit of 0.22~mJy at 3.5~cm was found on 18 October 2006, 
	whilst on 9 June 2007 a point-source flux of $0.63\pm0.06$~mJy detected, demonstrating that the emission varies by a factor of 3 
	within a year. The emission at 3.5~cm was found to be unpolarised down to the noise level on 9 June 2007.
	$3\sigma$ upper limits of 0.20 and 0.40~mJy were determined at 6.3~cm on these two dates.
	
\begin{figure}
		\centering
		\includegraphics[width=\columnwidth]{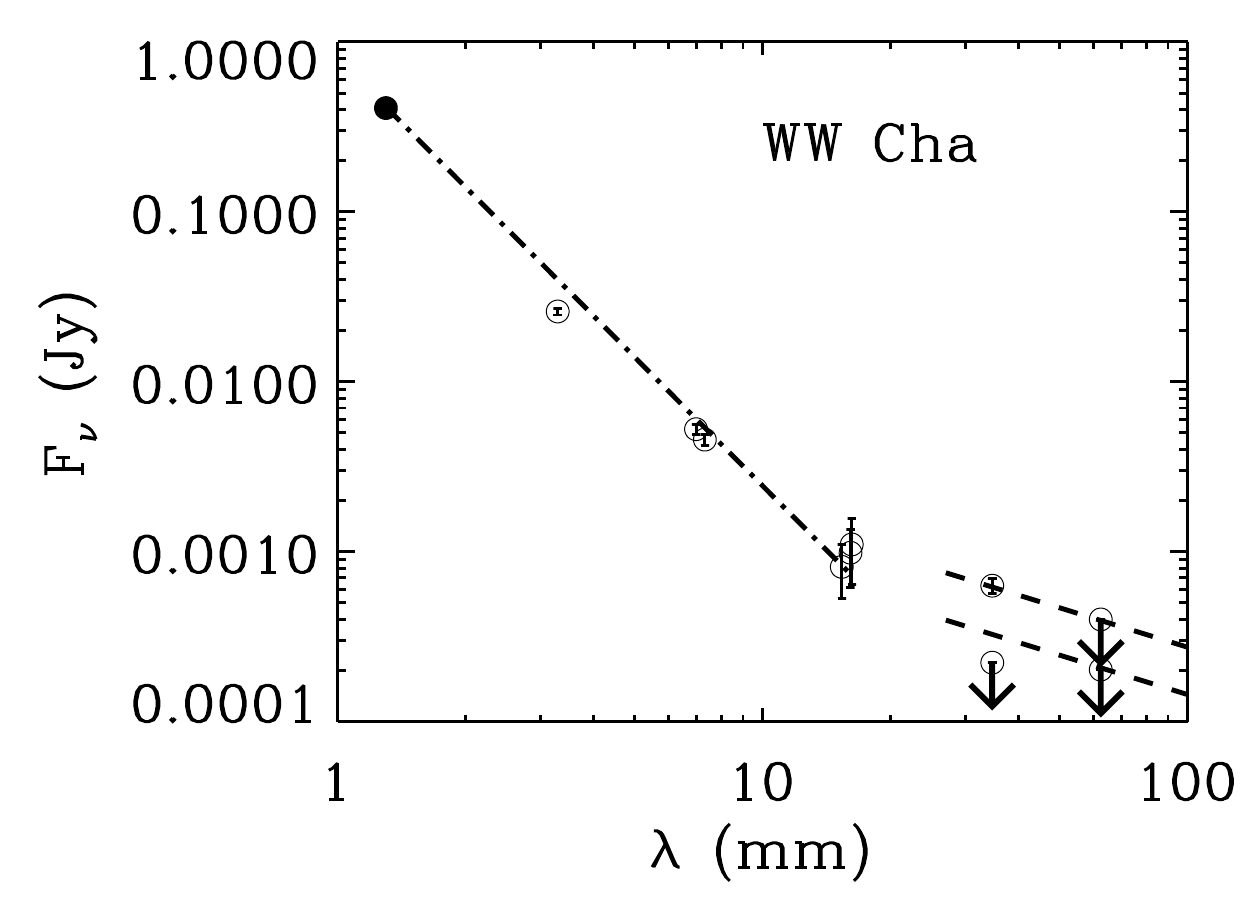}
		\caption[]{Long-wavelength SED of WW~Cha. The 1.3~mm point is from \citet{henning:1993}, the 3.3~mm point from
			\citet{lommen:2007}, and the other points are from this work [point-source fits in the ($u, v$) plane].
			The dash-dotted line shows the fit to the mm data points from 1.3 through 16.2~mm with $\alpha = 2.52$.
			The dashed lines show $\alpha_{cm} = 0.77$ slopes to the cm data. Note that the emission at 3.5~cm is
			variable, whereas the emission at 16~mm is stable to well within the uncertainties, indicating that the
			emission mechanisms at these wavelengths are different.}
		\label{fig: WWCha SED}
	\end{figure}

	In conclusion, WW~Cha is detected at 3, 7, 16~mm and 3.5~cm,  and upper limits were found at 6.3~cm, as depicted in 
 	Fig.~\ref{fig: WWCha SED}. 
	Most notably, the source was detected at 16~mm six out of 
	six times, and found stable over periods varying from days to years, while the 3.5~cm emission was found to vary. 
		
\subsubsection{Other sources in the field}

	At 16.2~mm, a second source was detected about 40$\arcsec$ to the north-west of WW~Cha. This source was identified 
	with the YSO Ced~112~IRS~4 -- see Fig.~\ref{fig: IRAC}. The data at 
	16.2~mm were combined in the ($u, v$) plane, yielding a point-source flux of $0.22 \pm 0.08$~mJy. The source was
	subsequently also detected at 7~mm with a point-source flux of $0.77 \pm 0.14$~mJy, and the source was not detected
	at 3~mm, down to a 3$\sigma$ upper limit of 3.3~mJy. This implies a rather shallow mm slope ($\alpha \approx 1.5$),
	possibly indicating a large contribution from, e.g., free-free emission at 16 and maybe also 7~mm.

\begin{figure}
		\centering
		\includegraphics[width=\columnwidth]{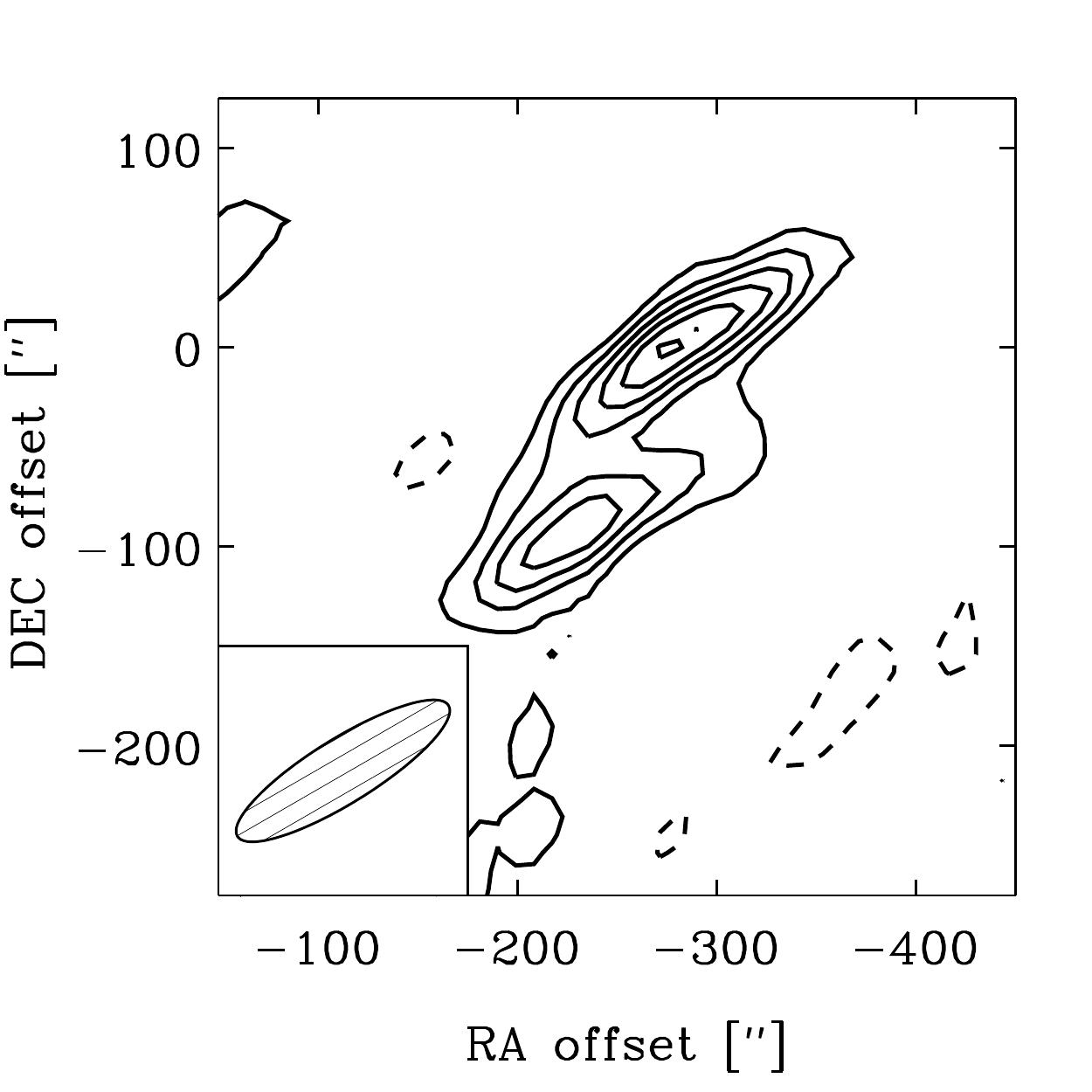}
		\caption[]{Image of the region 4~arcmin to the west of WW~Cha at 6.3~cm. The offsets are with respect to the
			phase centre, which is at 11:10:04.0, -76:35:59.0 (J2000). Contours are at 2, 4, 6, 8, 10, 12 times the rms
			of 0.19~mJy/beam; negative contours are dashed.}
		\label{fig: jets?}
	\end{figure}

	In addition, two large lobe-like features 
	were detected about $4\arcmin$ to the west of WW~Cha at both 3.5 and 6.3~cm. 
	These as yet unidentified radio sources are clearly seen in the 6~cm data, as shown in Fig.~\ref{fig: jets?}, which presents the 
	combined 6.3~cm data from 18 October 2006 and 9 June 2007.
	The positions of the sources are 11:08:43, -76:34:58 (J2000) (northern source) and 11:08:56, -76:36:24 (J2000) (southern source).
	The sources show up in both epochs at which WW~Cha was observed at 6.3~cm, but they were not both positively detected at 3.5~cm in each epoch
	(see Table~\ref{tab: results double quasar}). Using the results from point-source fits in the ($u, v$) plane, spectral slopes between 
	3.5 and 6.3~cm of $\alpha < -2.7$ and $\alpha = -3.0 \pm 0.2$ are found for the northern source and of
	$\alpha = -2.54 \pm 0.19$ and $\alpha < -2.3$ for the southern source, for  18 October 2006 and 9 June 2007
	respectively. Recall that $\alpha$ is defined as $F_\nu \propto \nu^{\, \alpha}$, and a negative value implies a flux that increases towards 
	longer wavelengths. Such steeply rising negative spectra are generally attributed to non-thermal emission. Neither of the two sources 
	were detected at 16~mm.  The sources may be background radio galaxies, though their fluxes are lower than the limiting magnitude of the 
	6~cm Parkes-MIT-NRAO radio continuum survey. 
	\begin{table*}
		\caption[]{Overview of the results for two radio sources detected to the west of WW~Cha in the 3 and 6~cm bands.  
		Point source fluxes  are given from both the northern and southern source.}\label{tab: results double quasar}
		\centering
		\begin{tabular}{lccccc}
			\hline
			\hline
			Obs. date	  	& Wavelength  & \multicolumn{2}{c}{Flux$^\mathrm{a}$}	& RMS$^\mathrm{b}$	& Beam size$^\mathrm{c}$	\\
					  	& (cm)	 	& \multicolumn{2}{c}{(mJy)}			& (mJy/beam)			& (arcsec)				\\
					  	&			& (northern)			& (southern)	&					&					        \\
			\hline
			18 October 2006 & 3.47		& $< 0.27$			& $0.29$		& 0.09				& $71 \times 20$			\\
			9 June 2007	   & 3.47		& 0.29  				& $< 0.25$    	& 0.08				& $53 \times 16$			\\ \hline	
			18 October 2006 & 6.25		& 1.34				& 1.29		& 0.16				& $125 \times 32$			\\ 
			9 June 2007	   & 6.25		& 1.71				& 0.99		& 0.14				& $91 \times 26$			\\	
			\hline
		\end{tabular}
		\begin{list}{}{}
			\item[$^\mathrm{a}$] Continuum fluxes from point-source fits in the ($u, v$) plane.  3$\sigma$
				upper limits are quoted in the case of non-detections.
			\item[$^\mathrm{b}$] RMS calculated from the cleaned image.
			\item[$^\mathrm{c}$] Restored beam, using natural weighting.
		\end{list}
	\end{table*}

\subsection{RU~Lup}

\begin{figure}
		\centering
		\includegraphics[width=\columnwidth]{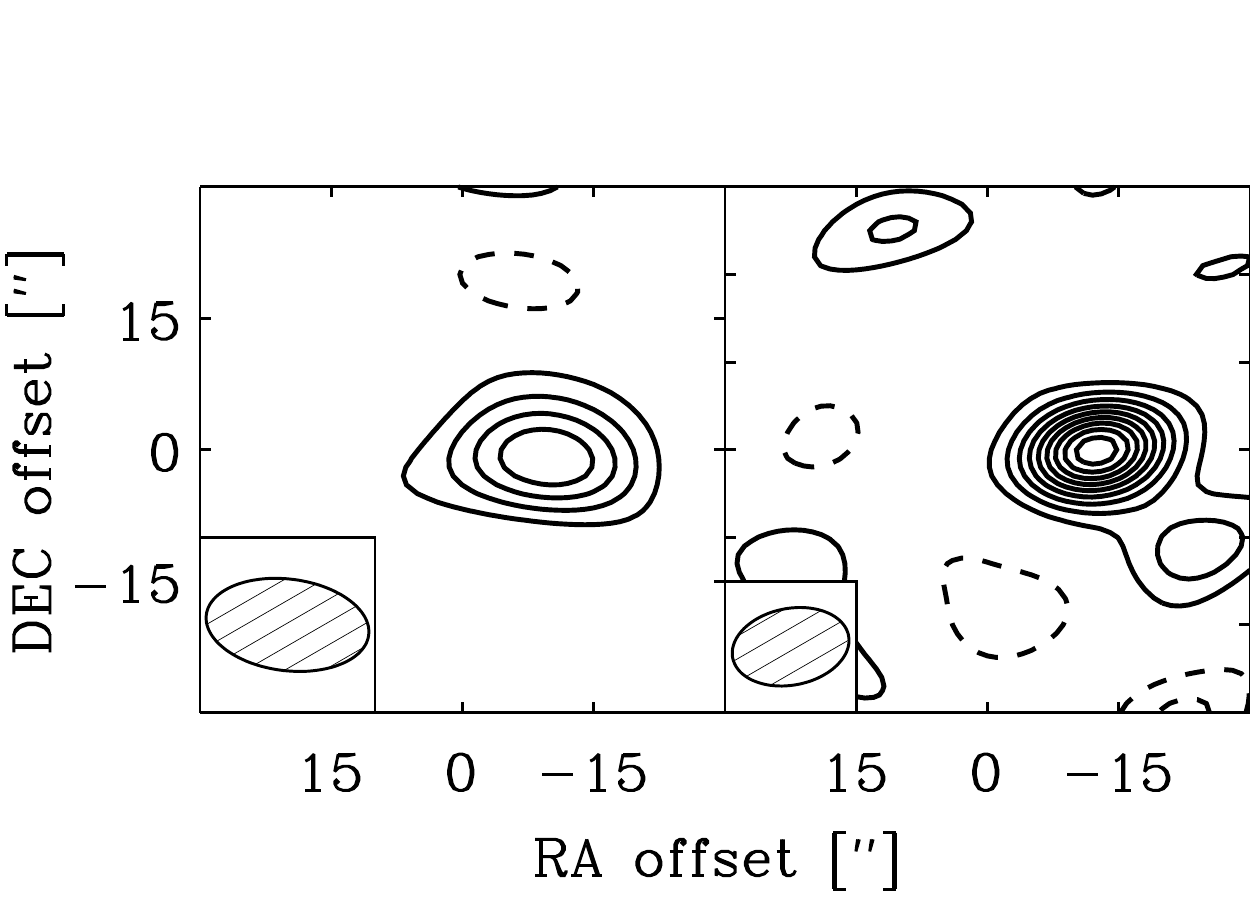}
		\caption[]{Image of RU~Lup, observed at 7.3~mm on 6 October 2007 (left panel) and at 16.2~mm on 11 October 2006
			(right panel). The offsets are with respect to the phase centre, which is located at 15:56:42.2, -37:49:15.5 (J2000).
			The contours are at 2, 4, 6, ... times the rms of 0.11~mJy/beam for both panels; negative contours are dashed. The peak
			emission corresponds to the position of the optically visible star. Note that the (4$\sigma$) peaks to the
			north-east and south-west of RU~Lup at 16.2~mm are probably not real, as they are equally ``significant''
			as the negative sidelobe in the lower right corner.}
		\label{fig: rulup at 7 and 16 mm}
	\end{figure}

	Table~\ref{tab: results RU Lup} summarises the fluxes of RU~Lup on the different dates at the various wavelengths.
	\begin{table*}
		\caption[]{Overview of the results for RU~Lup.}\label{tab: results RU Lup}
		\centering
		\begin{tabular}{lcccc}
			\hline
			\hline
			Obs. date		& Wavelength	& Flux$^\mathrm{a}$	&RMS$^\mathrm{a}$	& Beam size$^\mathrm{b}$	\\
						& (mm)		& (mJy)				& (mJy/beam)			& (arcsec)				\\	
			\hline
			\multicolumn{5}{c}{ATCA 7 mm band}																									\\
			\hline
			6 October 2007	& 7.0		& $1.83\pm0.25$	&$0.490$			& $18 \times 10$			\\
			6 October 2007	& 7.3		& $1.52\pm0.24$	& $0.141$			& $19 \times 10$			\\
			\hline
			\multicolumn{5}{c}{ATCA 12 mm band}																								\\
			\hline
			11 October 2006	& 16.1		& $2.28\pm0.09$	& $0.110$		 	& $14 \times 9$			\\
			11 October 2006	& 16.2		& $2.38\pm0.09$	& $0.113$			& $13 \times 9$			\\
			24 October 2007	& 15.4		& $<1.08$ 		& $0.360$			& $9.1 \times 7.8$		\\
			24 October 2007	& 16.2		& $<0.85$ 		& $0.283$			& $9.4 \times 8.1$		\\
			4 November 2007	& 15.4		& $<0.39$ 		& $0.131$			& $6.5 \times 1.5$	\\
			4 November 2007	& 16.2		& $0.51\pm0.11$	&$0.124$			& $6.7 \times 1.6$	\\
			\hline
			\multicolumn{5}{c}{ATCA 3 cm band}																									\\
			\hline
			12 October 2006	& 34.7		& $ <0.27$		& $0.089$			& $27 \times 18$			\\
			13 October 2006 	& 34.7		& $ 0.35\pm0.12 $	& $0.098$			& $23 \times 19$			\\
			9 June 2007		& 34.7		& $< 0.25$		& $0.082$			& $53 \times 19$			\\
			4 November 2007	& 34.7		& $< 0.49$		& $0.164$			& $56 \times 3$			\\
			\hline
			\multicolumn{5}{c}{ATCA 6 cm band}																									\\
			\hline
			12 October 2006	& 62.5		& $ 0.36\pm0.07$	& $0.111$			& $46 \times 34$			\\
			13 October 2006 	& 62.5		& $ 0.72\pm0.10$	& $0.128$			& $44 \times 33$			\\
			9 June 2007		& 62.5		& $0.40 \pm 0.07$	& $0.087$			& $91\times 31$			\\
			4 November 2007	& 62.5		& $<0.44$			& $0.145$			& $99 \times 6$			\\
			\hline
		\end{tabular}
		\begin{list}{}{}
			\item[$^\mathrm{a}$] Continuum fluxes are from point-source fits in the ($u, v$) plane, where 
				3$\sigma$ upper limits are quoted in the case of non-detections. RMS calculated from the cleaned image.
			\item[$^\mathrm{b}$] Restored beam, using natural weighting. Note that the exact beam size depends
				on frequency and location of the source in the sky, as well as on the time of the observations,
				over which time range the data were taken, and exactly which data/baselines were included.
		\end{list}
	\end{table*}
	The source RU~Lup was observed at 3.3~mm with the ATCA on 24 August 2005 and reported to have a point-source flux of 12.7~mJy \citep{lommen:2007}.
	RU~Lup was observed at 7~mm only once and clearly detected (see Fig.~\ref{fig: rulup at 7 and 16 mm}). Point-source fluxes of 
	$1.83\pm0.25$~mJy at 7.0~mm  and of $1.52\pm0.24$~mJy at 7.3~mm were obtained. 
	Though these values are consistent with each other, both wavelengths will be treated separately 
	again, for consistency.

	Three sets of observations were made of RU~Lup at 16~mm. It was detected on 11 October 2006 
	(see Fig.~\ref{fig: rulup at 7 and 16 mm}),
	with point-source fluxes of
	$2.28\pm0.09$ and $2.38\pm0.09$~mJy at 16.1 and 16.2~mm respectively. 	It was not detected 
	on 24 October 2007, with 3$\sigma$ upper limits of $1.08$ and $0.85$~mJy at 15.4 and 16.2~mm.
	A flux of $0.51\pm0.11$ at 16.2~mm 
	was detected on 4 November 2007, and a  3$\sigma$ upper limit of $0.39$~mJy at 15.4~mm, implying
	that the flux of RU~Lup dropped by at least a factor of six over the course of a year.
	The data were checked for polarisation on 11 October 2007, when the emission was strongest.
	However, the emission was found to be unpolarised down to the noise level of our observations.

\begin{figure}
		\centering
		\includegraphics[width=\columnwidth]{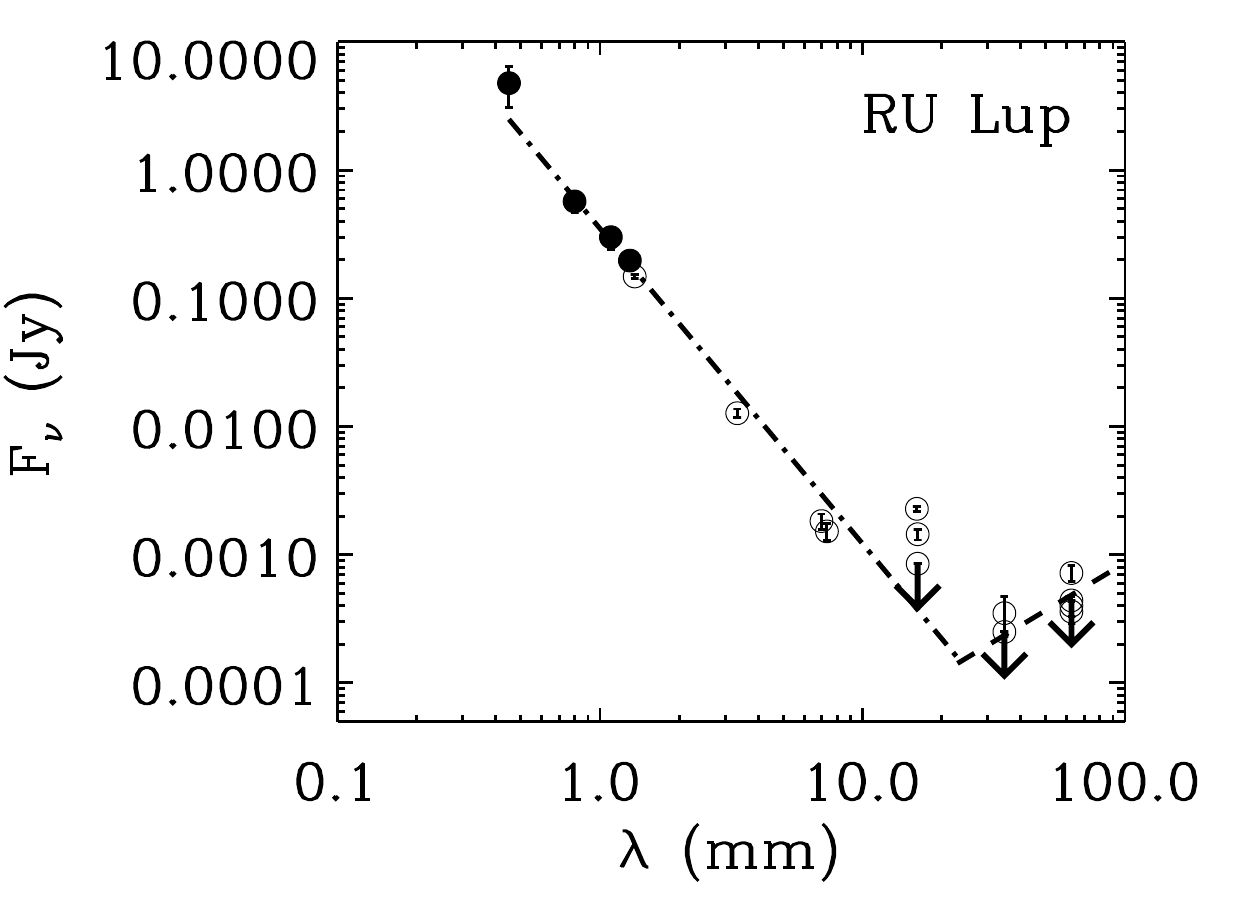}
		\caption[]{Long-wavelength SED of RU~Lup. The dash-dotted line shows a fit to data from 450~$\mu$m through 7.3~mm
			of $\alpha = 2.46$, using data from \citet{weintraub:1989}, \citet{nurnberger:1997}, \citet{lommen:2007},
			and this work. The dashed line shows an $\alpha_{cm} = -1.23$ slope to the cm data. Multiple points are shown
			at 16~mm, 3.5~cm, and 6.3~cm, to indicate the source's variability at those wavelengths.}
		\label{fig: RULup SED}
	\end{figure}

	At 3.5 and 6.3~cm, RU~Lup was observed on four different occasions (see Table~\ref{tab: results RU Lup}). It was detected 
	in most data sets, but the flux varied by up to a factor of two over the course of a year.
	Thus, RU~Lup was detected at wavelengths ranging from 3.3~mm to 6.3~cm, with variable emission at 16~mm and longer  
	wavelengths.  The SED is given in Fig.~\ref{fig: RULup SED}. 

\subsection{CS~Cha}

\begin{figure}
		\centering
		\includegraphics[width=\columnwidth]{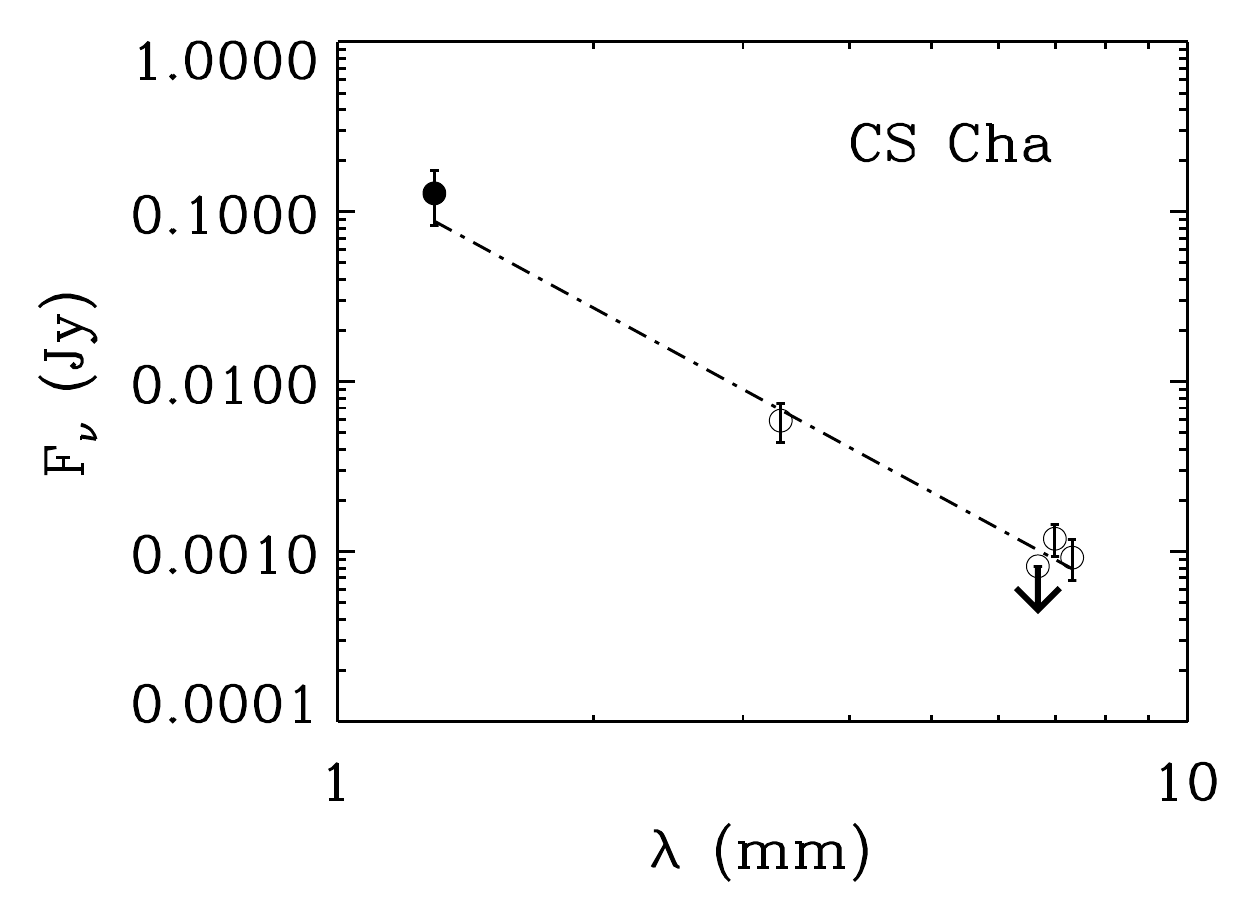}
		\caption[]{Millimetre SED of CS~Cha. The 1.3~mm point is from \citet{henning:1993}, the 3.3~mm point from \citet{lommen:2007},
			and the 7~mm points are from this work [point-source fits in the ($u, v$) plane]. The dash-dotted line shows the
			fit to the mm data points of $\alpha = 2.9$.}
		\label{fig: CSCha SED}
	\end{figure}

	Table~\ref{tab: results CS Cha} summarises the  7~mm fluxes of CS~Cha on the different dates. 
	CS~Cha was previously observed at 3.3~mm with the ATCA on 26 August 2005 and detected with a point-source flux of 5.9~mJy \citep{lommen:2007}. 
	The phases on 30 June 2008 were so unstable that the data could not be used in the analysis.
	The source was detected once at 7.3~mm, twice at 7.0~mm, and upper limits were determined at 6.7~mm. 
	No proper map could be extracted from the data, in the case of 26 April 2008 because most of the baselines had to be flagged as bad and
	no closure could be reached, and in the case of 6 July 2008 because of the very elongated beam due to the short observing time.
	The analysis was done in the ($u, v$) plane, and a point-source flux of 
	$0.92\pm0.25$~mJy was found at 7.3~mm on 26 April 2008. An average 7.0~mm point-source flux of $1.19\pm0.27$~mJy 
	(arithmetic mean of the fluxes and quadratic mean of the uncertainties) was determined from the two detections. 	 
	None of the 7~mm-band detections were resolved.  
	The SED of CS~Cha is shown in Fig.~\ref{fig: CSCha SED}.

	\begin{table*}
		\caption[]{Overview of the results for CS~Cha.}\label{tab: results CS Cha}
		\centering
		\begin{tabular}{lccccc}
			\hline
			\hline
			Obs. date		& Wavelength	& Flux$^\mathrm{a}$		& RMS$^\mathrm{a}$	& Beam size$^\mathrm{b}$	\\
						& (mm)		& (mJy)					& (mJy/beam)			& (arcsec)				\\
			\hline
			26 April 2008	& 7.0			& $1.00\pm0.28$		& 	0.129			& $2.4 \times 1.4$		\\
			26 April 2008	& 7.3			& $0.92\pm0.25$		& 	0.114			& $2.4 \times 1.5$		\\
			5 July 2008	& 6.7			& $< 0.82$			& 	0.273			& $10.8 \times 0.6$		\\
			5 July 2008	& 7.0			& $<0.71$				& 	0.238			& $11.3 \times 0.6$		\\
			6 July 2008	& 6.7			& $< 1.11$			& 	0.371			& $4.3 \times 0.6$		\\
			6 July 2008	& 7.0			& $1.38\pm0.26$		& 	0.218			& $4.5 \times 0.6$		\\
			\hline
		\end{tabular}
		\begin{list}{}{}
			\item[$^\mathrm{a}$] Continuum fluxes are from point-source fits in the ($u, v$) plane, where 
				3$\sigma$ upper limits are quoted in the case of non-detections. RMS calculated from the cleaned image.
				CS~Cha is located at 11:02:25.1, -77:33:35.95 (J2000).
			\item[$^\mathrm{b}$] Restored beam, using natural weighting. Note that the exact beam size depends
				on frequency and location of the source in the sky, as well as on the time of the observations,
				over which time range the data were taken, and exactly which data/baselines were included.
		\end{list}
	\end{table*}

\section{Discussion and interpretation}\label{sect: discussion}

The YSOs WW~Cha, RU~Lup, and CS~Cha were observed at 7~mm using the ATCA. 
WW~Cha and RU~Lup were furthermore observed at 16~mm, 3.5~cm, and 6.3~cm. 
In this Section we discuss the implications of our observations and in particular we try to determine the source of the
emission at cm wavelengths.

\subsection{WW~Cha}
 
The SED of WW~Cha (Fig.~\ref{fig: WWCha SED}) shows a break in slope at 16~mm. This suggests that the 16~mm 
flux is a continuation of the mm flux, implying that the emission up to that wavelength comes from thermal dust 
emission.  
The stability of the 16~mm flux over time scales of days, months and years (see Fig.~\ref{fig: 18496 MHz}) supports the 
hypothesis that it is due to thermal emission from large, cool dust grains, and makes it unlikely that it is due to 
stellar magnetic activity, which is known to show variations up to an order of magnitude or more over timescales 
varying from months to years \citep[e.g.][]{kutner:1986, chiang:1996}.

\subsubsection{Disk emission}
 
If the mm emission (from 3 to 16~mm) is due to dust, we should be able to resolve the emission (on the scale of the 
disk).  To determine whether the source is resolved at any of these wavelengths, the method of \citet{lommen:2007} was 
followed. A circular Gaussian was fitted in the ($u, v$) plane and if the Gaussian fitted flux is at least 2$\sigma$ 
larger than the flux from a point-source fit, the source is concluded to be resolved at that wavelength.  Using this 
definition, the 7~mm emission was found to be resolved on 31 March 2008 and the 16~mm emission was resolved on 2 
November 2007 when WW~Cha was observed with the 1.5A array configuration. Note that these two dates have the 
lowest RMS in their respective wave bands, as well as the smallest beam sizes. For details, see 
Table~\ref{tab:resolvingWWCha} (and note that the 2 November 2007 data include antenna 6).
Fitting a circular Gaussian in the ($u, v$) plane to the emission of WW~Cha at 3~mm gives a flux of 33.1~mJy 
\citep{lommen:2007} and yields a size of $1.5 \pm 0.2$~arcsec, corresponding to a physical (disk) size of 
$240 \pm 30$~AU at a distance of 160~pc \citep{whittet:1997}. 
A circular Gaussian fitted to the 7.0~mm emission in the ($u,v$) plane 
gives a Gaussian flux of $5.79\pm0.38$~mJy and a corresponding source size of $2.0$~arcsec, consistent with the size at 
3~mm.   
Similarly, for the 15.4~mm emission a Gaussian flux of $1.65\pm0.55$~mJy and a corresponding source size of 
$2.1$~arcsec are obtained, which is consistent with both the 3 and 7~mm sizes.  See Table~\ref{tab:resolvingWWCha} for 
full details.
	\begin{table*}
		\caption[]{Resolving WW~Cha. Point-source versus Gaussian fluxes and resulting Gaussian source size for 
		WW~Cha at 7 and 16~mm.}\label{tab:resolvingWWCha}
		\centering
		\begin{tabular}{lcccccccc}
		\hline\hline
		Obs. date			& Wavelength	& Flux (p)$^{\rm a}$	& Flux (G)$^{\rm a}$	& RMS$^{\rm b}$		& Gaussian size	& Beam size$^{\rm c}$	& Dust opacity$^{\rm d}$	& Disk mass$^{\rm e}$	\\
						& (mm)		& (mJy)			& (mJy)			&(mJy/beam)		& (arcsec)	&(arcsec)		& (cm$^2$~g$^{-1}$)		& (M$_\odot$)		\\
		\hline
 		24-28 August 2005$^{\rm f}$	& 3.3		& $25.9\pm 1.0$		&$33.1\pm 2.1$		&$1.2$			&$1.32$		&$2.5\times2.2$	 	& 2.86				& 0.024			\\ 
		31 March 2008			& 7.0		& $5.1\pm0.19$		&$5.79\pm0.38$		&$0.231$		&$2.0$		&$5.8\times5.3$	 	& 1.94				& 0.027			\\ 
		31 March 2008			& 7.3		& $5.2\pm0.17$		&$6.46\pm0.38$		&$0.194$		&$3.4$		&$6.1\times5.6$	 	& 1.90				& 0.033			\\ 
	        2 November 2007			&15.4		&$0.55\pm0.20$		&$1.65\pm0.55$		&$0.125$		&$2.1$		&$1.9\times1.3$ 	& 1.29				& 0.055			\\
	        2 November 2007			&16.2		&$0.48\pm0.12$		&$1.20\pm0.26$		&$0.089$		&$1.2$		&$2.0\times1.4$ 	& 1.26				& 0.045			\\
		\hline
		\end{tabular}
		\begin{list}{}{}
			\item[$^\mathrm{a}$] Continuum fluxes are from point-source (p) and circular Gaussian (G) fits in the ($u, v$) plane.
			\item[$^\mathrm{b}$] RMS calculated from the cleaned image.
			\item[$^\mathrm{c}$] Restored beam, using natural weighting. 
			\item[$^\mathrm{d}$] Using the opacity law of \citet{beckwith:1990} $\kappa_\nu$ = 10 ($\nu/10^{12}$~Hz)$^\beta$, with $\beta = 0.52$.
			\item[$^\mathrm{e}$] Assuming a gas-to-dust ratio of 100.
			\item[$^\mathrm{f}$] From \citet{lommen:2007}.
		\end{list}
	\end{table*}

 	The peak brightness temperature of the emission is given by
	\begin{equation}
		T_b = \frac{F_\nu c^2}{2 \nu^2 k} \frac{1}{\theta^2},
	\end{equation}
	where $k$ is Boltzmann's constant and $\theta^2$ the area of the emitting source. 
	The 16.2~mm flux of $1.20 \pm 0.32$~mJy, (see Table~\ref{tab:resolvingWWCha}, but now including a 15\% 
	uncertainty from the absolute flux calibration), implies a brightness temperature $T_b = 4.3 \pm 1.2$~K. The 
	low brightness temperature rules out that the emission is thermal free-free from an ionised gas 
	\citep[e.g.][]{morgan:1990, snell:1986}, unless (a) the emission has very low optical depth, or (b) it has a 
	very low filling factor within the apparently resolved region (or a combination of these).
	Finally, a fit to the SED of WW~Cha (Fig.~\ref{fig: WWCha SED}) from 1.3 to 16.2~mm gives 
	$\alpha=2.52\pm0.12$. Given the stability of the 16~mm flux, the resolved size of the emission, the low peak 
	brightness temperature, and the slope of the SED, we conclude that the emission at 16~mm is thermal emission 
	from large grains or small ``pebbles''.
	
	From our previous 3.3~mm observations, \citet{lommen:2007} determined the disk mass for WW~Cha as
	0.077~M$_\odot$. However, this mass determination assumed a dust opacity at 3.3~mm of
	$\kappa_\nu = 0.9$~cm$^2$~g$^{-1}$. Given the shallow slope of the SED in the mm regime for WW~Cha, this value
	appears to be rather small. Indeed, if we take the mm slope $\alpha=2.52\pm0.12$ and assume that the 
	contribution of optically-thick emission is negligible at these wavelengths, we find 
	$\beta \approx \alpha - 2 = 0.52$. Using the opacity law of \citet{beckwith:1990}\footnote{Note that 
	\citet{beckwith:1990} estimate $\kappa_\nu = 0.1 (\nu/10^{12}$~Hz$)^\beta$ cm$^2$ g$^{-1}$, where 
	$\kappa_\nu$ is the opacity index {\it for the gas and the dust combined}, i.e., with an implied gas-to-dust ratio. 
	Our values for $\kappa_\nu$, however, are for the dust alone, and hence we have to account for the gas-to-dust ratio
	explicitly in the calculation of the disk mass.}, 
	$\kappa_\nu = 10 (\nu/10^{12} {\rm Hz})^\beta$~cm$^2$~g$^{-1}$, this yields a dust opacity at 3.3~mm of
	$\kappa_\nu = 2.86$~cm$^2$~g$^{-1}$ and consequently a disk mass of $M_{\rm disk} = 0.024$~M$_\odot$, where we use
		\begin{equation}
		\label{eq: disk mass}
		M_{\rm disk} = \frac{F_\nu \Psi D^2}{\kappa_\nu B_\nu(T_{\rm dust})},
	\end{equation}
	with $\Psi$ the gas-to-dust ratio (taken to be 100), $D$ the distance to the source, $\kappa_\nu$ the dust 
	opacity, and $B_\nu(T_{\rm dust})$ the brightness at the frequency $\nu$ for a dust temperature $T_{\rm dust}$
	(taken to be 25~K), as given by the Planck function. Disk masses in the range of
	0.024 to 0.055~M$_\odot$ are found for our observations out to 16~mm using this opacity law (see 
	Table~\ref{tab:resolvingWWCha}). The spread in disk masses can in part be explained by the uncertainty in the 
	values of the fluxes, and is also an indication that the opacity law is not entirely applicable to the case at hand.
	However, a disk mass of $\sim$0.03~M$_\odot$ for the system is probably accurate to within a factor of a
	few. Taking into account the contribution of optically-thick emission, i.e., $\Delta \approx 0.2$ in 
	Eq.~\ref{eq: beta} \citep{lommen:2007}, does not significantly change
	the results, giving disk masses in the range of 0.03 to 0.07~M$_\odot$. 
	%
	A more detailed determination of the disk mass requires radiative-transfer modelling of the system, including
	the full SED and available spatial information. For this, we refer the reader to an up coming paper
	(M\'{e}nard, et al., in prep.).

\subsubsection{Emission at longer wavelengths}

	The slope of the 3.5 to 6.3~cm part of the SED gives $\alpha_{\rm cm} > 0.77$, which 
	suggests that the emission is thermal, perhaps from an optically-thick, ionised wind 
	\citep{panagia:1975, wright:1975, olnon:1975}. However, at 3.5~cm WW~Cha appears to change in flux by a
	factor of three between the two epochs at which it was observed (18 October 2006 and 9 June 2007). This amount 
	of variability makes optically-thick wind emission an unlikely source for the cm emission. 
	Indeed it is difficult to determine a spectral index in the cm, as the 3.5~cm flux clearly varies and only 
	upper limits were obtained at 6.3~cm. \citet{smith:2003} found that for T~Tau the cm spectral index can be 
	positive even for non-thermal emission. They attribute the highly variable emission at 3.5~cm, along with 
	a high level of polarisation and a very compact emission region, to an origin in a magnetically-dominated
	region close to the star. However, this is unlikely to be the cause of the emission at 3.5~cm for WW~Cha, which we
	found to be unpolarised down to the noise level of our observations. Without information about the size of 
	the emitting region, it is difficult to determine the nature of the cm emission in  
	WW~Cha. Follow-up observations should include very-long-baseline-array observations, which would resolve
	the system down to scales of several stellar radii, such as has been done for, e.g., the double binary 
	T~Tau \citep{smith:2003} and for the binary system V773 Tauri A \citep{massi:2008}. We can 
	conclude, however, that the emission at cm wavelengths
	is not dominated by thermal emission from large, cm-sized grains.

\subsection{RU~Lup}

	RU~Lup was detected at 3.3, 7.0, and 7.3~mm. It was also detected at longer wavelengths, although not every time
	the source was observed. The
	SED shows a break around 16~mm (Fig.~\ref{fig: RULup SED}), indicating that the dominating emission mechnism
	changes, as was also observed for WW~Cha.

\subsubsection{Disk emission}
 
	\citet{lommen:2007} found the RU~Lup to be resolved at 1.4 and 3.3~mm, with 
	sizes of $1.02 \pm 0.32$ and $0.99 \pm 0.32\arcsec$ respectively, and a 
	(sub)mm slope $\alpha = 2.5 \pm 0.1$ from 450~$\mu$m through 3.3~mm, indicating that the mm emission is due to grains 
	of at least mm sizes.  Adding the data points at 7.0 and 7.3~mm to the SED of RU~Lup (see Fig.~\ref{fig: RULup SED}), a 
	slope of $\alpha = 2.46 \pm 0.09$  from 450~$\mu$m through 7.3~mm is found, indicating
	that the emission at 7~mm is still due to dust.  Fitting a circular Gaussian in the ($u,v$) plane gives a Gaussian flux of 
	$1.64\pm0.39$ and $2.33\pm0.39$~mJy at 7.0 and 7.3~mm respectively. This suggests that the 7.3~mm data are resolved 
	(since $F_{\rm Gauss} > F_{\rm point} + 2\sigma$) with a Gaussian size of $9.3\arcsec$, which is about 1300~AU at 140~pc.
	This is quite large, suggesting an extended envelope rather than a disk. This Gaussian size gives a very low peak brightness 
	temperature ($<0.1$~K). However, the evidence for extended emission is borderline, with the 7.0~mm flux failing our ``resolved'' 
	test $F_{\rm Gauss} > F_{\rm point} + 2\sigma$ and the 7.3~mm emission only just meeting this criterion.

	\citet{lommen:2007} estimated the disk in the system to be 0.032~M$_\odot$ in mass, assuming an opacity at 3.3~mm of 
	0.9~cm$^2$~g$^{-1}$. Again neglecting the contribution from optically-thick emission, an opacity index $\beta \approx \alpha - 2 = 0.46$ is 
	found. Using the opacity law of \citet{beckwith:1990}, this implies an opacity of 3.30, 2.35, and 2.30~cm$^2$~g$^{-1}$ at 3.3, 7.0, 
	and 7.3~mm, respectively. This in turn yields a disk mass of 0.010, 0.006, and 0.010~M$_\odot$ for the different wavelengths, where 
	the spread is most likely dominated by the uncertainty in the fluxes. Again, we refer the reader to an upcoming paper (M\'{e}nard, 
	et al., in prep.) for a more detailed calculation of the disk mass.

\subsubsection{Emission at longer wavelengths}

	RU~Lup was detected twice at 16~mm, with a $3\sigma$ upper limit found from a third observation. The fluxes from the three 
	different data sets are inconsistent with each other (see Fig.~\ref{fig: 18496 MHz}), ruling out thermal emission from large grains 
	at these wavelengths. 
	The centimetre slope of the SED from 3.5 to 6.3~cm is $\alpha_{\rm cm} = -1.23\pm0.17$ (taken from the detections on 13 October 2006),
	which, in contrast to WW~Cha, implies non-thermal emission from	optically thin gyrosynchrotron emission \citep{anglada98,forbrich06}, though the variability 
	of the emission makes it difficult to fit one consistent slope to the data. 
	The  16~mm flux is surprisingly high (see Fig.~\ref{fig: RULup SED}) and  appears to 
	result from more than just thermal emission from dust grains. One possible explanation for the excess 16~mm flux and its variability is 
	that some of the emission comes from a cyclotron maser \citep{dulk:1985}. 
	The data from 11 October 2006, when the 16~mm flux was highest,  have negligible Stokes $Q$, $U$, and $V$ fluxes
	and hence the emission was unpolarised. Furthermore, maser cyclotrons are expected to produce emission at and around 
	the cyclotron frequency alone. This could result in the emission only being significantly stronger in one of the two sidebands,
	which is not observed. The lack of polarisation and strongly coherent emission seems to suggest that the emission is not due to a cyclotron
	maser. However, the overall frequency range of maser emission from a source could be larger than just directly around the cyclotron 
	frequency, and the intrinsic polarisation is likely
	to be destroyed by Faraday rotation in the overlying plasma \citep{dulk:1985}. Therefore, cyclotron maser emission cannot be fully ruled out.
	Information on different, even shorter timescales of the radio emission may shed additional light on the processes at play, and the
	increased sensitivity of the ATCA with the recent addition of the Compact Array Broadband Backend can be a useful tool in this.
	Unfortunately, RU~Lup was never observed at 16~mm, 3.5~cm and 6.3~cm on the same day and we know that the 3.5~cm varies 
	by a factor of two within 24 hours.  It seems likely that there are  three different emission mechanisms -- disk$+$star$+$wind -- acting 
	over these three cm wavelength bands.  This is different from WW~Cha, where dust emission dominates up to 16~mm.
	To determine of the emission mechanism at longer wavelengths requires the source to be resolved, which needs 
	very-long-baseline-interferometer observations.

\subsection{CS~Cha}

During the three months that over which CS~Cha was observed, the 7~mm emission was relatively stable and a mm slope $\alpha=2.90\pm0.26$ from 1.3 to 7.3~mm was found.  
This is consistent with the SED slope found by \citet{lommen:2007} from 1.3 to 3.3~mm ($\alpha=2.9\pm0.5$). Since neither the 3 or 7~mm 
emission was resolved, it cannot be ruled out that the mm emission is optically thick. 
However, if we assume the emission to be completely optically thin, $\alpha = 2.90$ implies a dust opacity index $\beta = 0.90$. This would suggest that the 7~mm emission from
CS~Cha is also predominantly due to thermal dust emission from mm-sized grains.

The disk mass was estimated by \citet{lommen:2007} to be 0.021~M$_\odot$. Taking the opacity law of \citet{beckwith:1990} and an opacity index $\beta = 0.90$, a mass of
0.016, 0.018, 0.016~M$_\odot$ is found at the different wavelengths, which is consistent with the value of \citet{lommen:2007} given the uncertainties in the fluxes. 
\citet{espaillat:2007a} found a disk mass of 0.04~M$_\odot$ through more detailed modelling, slightly larger than the value found with a simple opacity
law.

\subsection{Large grains in protoplanetary disks}

The mm slopes of WW~Cha, RU~Lup and CS~Cha are given by $\alpha=2.52\pm0.12, 2.46\pm0.09$, and $2.90\pm0.26$ 
respectively. This suggests that all three sources have grains of at least mm sizes, although it should be noted that the 7~mm emission of 
CS~Cha was not resolved, and hence the emission could be optically thick.
This suggests that grains grow from sub-micron sizes to at least mm sizes throughout the bulk of the disk within a few hundred thousand years.  
To date only one T~Tauri star has been found with large, cm-sized grains -- TW~Hya.   This was the only other source that has been monitored at cm wavelengths for long periods of time 
and was found to have stable thermal dust emission at 3.5~cm.  But is TW~Hya unique?  Our new 16~mm results show that the disk of WW~Cha 
also contains cm-sized grains (and we have similar results for the Herbig Ae star HD~100546; Wright et al. -- in preparation).  Furthermore, about 10\% of the 
16~mm emission from RU~Lup is likely to derive from pebbles as well, so it would appear that cm-sized grains in protoplanetary disks are not as rare as might have been 
expected.

It has been a long-standing problem in planet formation theory that boulders of about a metre in size fall into the central star before accumulating and growing to 
kilometre-sized planetesimals \citep{weidenschilling:1977}. Recent numerical simulations have shown that this so-called "metre-size barrier" can be overcome.
For example, \citet{johansen:2007} find that grain growth via gravitational collapse can be very efficient in the mid-plane of turbulent disks where streaming instabilities 
help concentrate grains that grow to several 100 km in a few thousand years. \citet{lyra:2008} show that grains trapped in Rossby waves excited at the edge of dead zones can 
grow to Mars-sized embryos in a few thousand years. \citet{brauer:2008} also find that grains grow rapidly in the near-laminar dead zone of disks and that the associated 
pressure maxima near evaporation fronts ensure that the newly formed boulders to not migrate radially.  These simulations all start with grains that are already at least 
cm-sized. Our results demonstrate that this is reasonable, and that with seeds of this size the metre-sized barrier can be overcome to produce fully-fledged protoplanets.

Ideally one would like to know the timescales of grain growth and so may be tempted to try to find an evolutionary sequence of grain 
size with age. For example, RU~Lup and  WW~Cha are both about 0.5~Myr and show grain sizes up to about a cm, while 
TW~Hya is 9~Myr and shows grains up to 3.5~cm.  However, it should be noted that ages of T~Tauri stars are notoriously difficult to 
determine, but more importantly, the data tell us that other, possibly different, emission mechanisms 
are stronger for WW~Cha and RU~Lup than for TW~Hya -- which may be 
age-related.    Clearly, a much larger sample is needed to draw any significant conclusions regarding grain growth up to pebble or even 
boulder sizes.

\section{Conclusions}\label{sec:concs}

We have been monitoring the mm and cm emission of two T~Tauri stars, WW~Cha and RU~Lup, over the course of several years with the ATCA, and more recently 
performed 7~mm observations of the third young T~Tauri star CS~Cha. We find that emission up to 7~mm for all three sources is well explained by thermal dust emission 
from mm-sized grains.  The stability of the 16~mm flux in WW~Cha, along with the low peak brightness temperature at this wavelength, indicates that this emission is 
dominated by even larger, cm-size ``pebbles", making it the second protoplanetary disk known to contain such large grains.  The 16~mm emission of RU~Lup may also include 
dust emission from pebbles, but other emission mechanisms appear to dominate at this wavelength.  

This work underlines the necessity to observe young stellar objects at multiple wavelengths and to monitor them over
extended periods of time, in order to disentangle the various candidate emission mechanisms at mm and cm wavelengths.
The ATCA is well suited to do this and with the upgrade of the correlator that is currently 
underway, increasing the bandwidth of the telescope by a factor of 16, extended surveys of southern protoplanetary 
disks  will soon be within reach. This will allow us to put more stringent constraints on the processes involved in the first steps of
planet formation, telling us where and when they take place.
	
\begin{acknowledgements}
	DL acknowledges Swinburne University and UNSW@ADFA for their hospitality. This work was partially supported by a 
	Netherlands Research School For Astronomy network 2 grant and a Netherlands Organisation for Scientific Research Spinoza
	grant (DL and EFvD),  a Swinburne Researcher Development Scheme, a Swinburne Special Studies Program, and the
	Programme National de Physique Stellaire (PNPS), INSU/CNRS (STM), and by an ARC 
	Australian Research Fellowship (CMW). We would like to thank the ATNF staff at Narrabri for their hospitality and assistance, and 
	Annie Hughes and Steve Longmore for assisting with some of the observations. We also thank Fran\c{c}ois M\'{e}nard for useful 
	discussions, and we are indepted to the anonymous referee, whose comments helped to considerably improve this paper. This research 
	has made use of the SIMBAD database, operated at CDS, Strasbourg, France.
\end{acknowledgements}

\bibliographystyle{aa}
\bibliography{references}

\end{document}